\documentclass[preprint,aps,prd,showpacs]{revtex4}
\usepackage{revsymb}
\begin{document}
\renewcommand{\thesection}{\arabic{section}}
\renewcommand{\thesubsection}{\arabic{subsection}}
\title{Wien's Displacement Law in Rindler Space}
\author{Sanchari De$^{a.1)}$ and Somenath Chakrabarty$^{a,2)}$}
\affiliation{
$^{a)}$Department of Physics, Visva-Bharati, Santiniketan 731 235, 
West Bengal, India\\ 
$^1$E-mail: sancharide19@gmail.com\\
$^2$E-mail:somenath.chakrabarty@visva-bharati.ac.in}
\begin{abstract}
In this article we have developed the formalisms for the modified form of Wien displacement laws
for both the gas of electromagnetic waves and a gas of de Broglie
waves in Rindler space. In the case of de Broglie 
waves we assume both fermion type and boson type materials. Following the classic work of Wien, we assume 
that the wall of the enclosure containing the photon gas or the gas of de Broglie waves,
is expanding adiabatically with a uniform acceleration. 
\end{abstract}
\maketitle
\section{Introduction}
According to the principle of equivalence a frame undergoing uniform
accelerated motion is equivalent to a rest frame in presence of a
constant gravitational field \cite{R1,R2,R3}. 
Therefore considering that the surface of the
enclosure which is filled up with
either a photon gas or a gas of de Broglie waves is expanding 
adiabatically with  uniform  acceleration may be assumed to be
equivalent to a uniformly accelerated reference frame. Then following the principle of equivalence
we may replace it by a rest frame in
presence of a constant gravitational field. We further assume that
the gravitational field is uniform within a width $\Delta H$ inside
the enclosure near the
surface. Further, assuming that the structure of the enclosure is of
spherical in nature,
let $H$ be the corresponding radius.
It is quite obvious that a photon will gain or loose
energy when it travels towards the wall or reflected back from the
wall respectively \cite{R4,R5}. Hence we can say that the
picture will be reversed during the reflection from the wall (see also \cite{R55} for displacement law in the case
of particle production).
Assuming that the wall is
perfectly reflecting, then there will be no absorption of photon energy.
 the photon will be Doppler shifted both
during its incidence and reflection at the wall. However, since the
wall is expanding with uniform acceleration, unlike 
the case of expansion of the wall with a uniform velocity, in the
present scenario
the change is not so called apparent. Here the Doppler shift is 
associated with the
real change in energy of the photons. It is therefore the gravitational
redshift, not the conventional change in wavelength which occurs when
there is a relative motion between the source and the observer.
The wave will be blue shifted while moving towards the surface and
the reverse is the case, i.e., redshift of the wave during
reflection. We will show that the same is true for the matter wave
associated with the incident and the reflected particles (fermions or
bosons) when the enclosure is filled with either Fermi gas or Bose
gas.

To get a physical
insight of the incidence and the reflection processes, we consider
the prescription of Feynman instead of considering Rindler
space-time coordinate transformations \cite{R1,R3,R6} 
(see also \cite{R61,R62,R63}). 
These two approaches however, essentially give the
same result. Moreover in our opinion the prescription of Feynman lecture is
more straight forward and is easy to understand physically. Now the Rindler
coordinate transformations are exactly like the Lorentz
transformations in a frame undergoing uniform acceleration, otherwise
in flat Minkowski space-time geometry. 

In this article we have also obtained the 
modified form of Wien's displacement laws for de Broglie waves. The geometrical structure of the 
enclosure is exactly identical as we have considered in the case of photon gas.

In this article we have also studied the
gravitational redshift of photons using the prescription of Friedman and
Friedman et al \cite{R7,R8,R9,R91}. The later
approach is conventionally known as the extended 
relativistic dynamics with an upper
limit of acceleration. In this study we have also tried to put forward 
some physical meaning to the maximum value of acceleration.

The article is organized in the following manner. In the next
section,
following the prescription of Feynman lecture, we have obtained the
Doppler shift of photons and establish the modified form of Wien's
displacement law following Saha and Srivastava \cite{R10} when the wall of the enclosure is
expanding adiabatically with uniform acceleration.
In section-3 we have
obtained the gravitational redshift factor $z$ following Friedman and
Friedman et al \cite{R7,R8,R9,R91}. In section-4 we have developed the Wien's
displacement laws for the de Broglie waves for both fermion type and
boson type particles. Finally we have given 
conclusion of the work.
\section{Wien's Displacement Law in Rindler Space for Photon Gas}
Let us consider a photon of circular frequency $\omega_0$, traveling
towards the wall of the enclosure which is moving outward with a uniform
acceleration $\alpha$. The energy of the incident photon is then
$\varepsilon_0=\hbar \omega_0$. The equivalent moving mass is
therefore $m_0=\varepsilon_0/c^2=\hbar \omega_0/c^2$. Hence the
gain of energy by the photon is $m_0 \alpha \Delta H$, $\Delta H$ is
the distance traversed and $\alpha$ is assumed to be constant within
$\Delta H$ in the sense of principle of equivalence. Then the energy
of the photon at the instant of incidence on the wall is 
\begin{equation}
\varepsilon_1=\hbar \omega_0 +\frac{\hbar \omega_0}{c^2}\alpha \Delta
H
\end{equation}
Then the changed value of photon frequency is given by
\begin{equation}
\omega_1=\omega_0\left ( 1+\frac{\alpha \Delta H}{c^2}\right )
\end{equation}
This is the gravitational Doppler shifted frequency of the incident photon. The
same result can also be obtained from Rindler space-time coordinate
transformations. While
reflected back from the wall, the photon is traveling away from the
wall. The photon will therefore loose energy. If $\omega_2$ is the
Doppler shifted frequency of the reflected photon, then the energy of
this photon is given by 
\begin{equation}
\hbar \omega_1=\hbar \omega_2 -\frac{\hbar \omega_2}{c^2} \Delta H
\alpha 
\end{equation}
Hence 
\begin{equation}
\omega_1=\omega_2\left ( 1-\frac{\alpha \Delta H}{c^2}\right )
\end{equation}
Therefore during incidence it is gravitational blue shift, whereas during reflection it is gravitational redshift.
Then combining eqns.(2) and (4), we have
\begin{equation}
\omega_2=\omega_0 \frac{\left (1+\frac{\alpha \Delta H}{c^2} \right
)}
{\left ( 1-\frac{\alpha \Delta H}{c^2} \right )}
\end{equation}
Therefore it is not the same kind of physical picture as discussed in
the book by Saha and Srivastava \cite{R10}. 
In the later case the wall is moving with
uniform velocity and the wave gets red-shifted both during incidence
as well as during reflection and we know from the knowledge of optics that
the change is apparent. Whereas in the case of accelerated frame of
reference the energy of the photon changes both during incidence and
reflection. The change of energy is proportional to the constant
gravitational field. 
Therefore although the final result gives redshift of the
photon, which is equivalent to gravitational redshift, but 
it is quite different from the conventional redshift of photons in
optics. Since either during incidence or reflection
some change in energy of the photon is occurring and it depends on the
strength of constant gravitational field, the change
is real in nature.

Now replacing $\omega_0=\omega=2\pi \nu$ and $\omega_2=\omega+\Delta
\omega= 2\pi (\nu +\Delta \nu)$, we have
\begin{equation}
\nu +\Delta \nu =\nu \frac{\left ( 1+\frac{\alpha \Delta
H}{c^2}\right ) }
{\left ( 1-\frac{\alpha \Delta H}{c^2}\right )}
\end{equation}
Assuming that the factor $\alpha \Delta H/c^2 \ll 1$, we have 
\begin{equation}
\frac{\Delta \nu}{\nu} \approx \frac{2\alpha \Delta H}{c^2}
\end{equation}
Hence we can also write 
\begin{equation}
\frac{\Delta \lambda}{\lambda} = -\frac{2\alpha \Delta H}{c^2}
\end{equation}
Now the single particle classical Hamiltonian is given by
\begin{equation}
H_0=\left ( 1+\frac{\alpha\Delta H}{c^2}\right )pc=L(\Delta H)pc
=\varepsilon
\end{equation}
where $L(\Delta H)$ is a function of $\Delta H$. Hence following the
standard result of statistical mechanics, the energy density may be
written as \cite{R11}
\begin{equation}
\epsilon =\frac{{\rm{constant}}~T^4}{L^3(\Delta H)}
\end{equation}
For adiabatic expansion of the photon gas
\begin{equation}
PV^{4/3}={\rm{constant}}
\end{equation}
Again for the photon gas
\begin{equation}
P=\frac{1}{3} \epsilon
\end{equation}
Hence
\begin{equation}
\epsilon V^{4/3}={\rm{constant}}
\end{equation}
The above equation can also be expressed as
\begin{equation}
\frac{T^4H^4}{\left (1+\frac{\alpha \Delta H}{c^2}\right )^3}=
{\rm{constant}}
\end{equation}
where we have used $V=4\pi H^3/3$, the volume of the enclosure.
Hence assuming that the factor $\alpha \Delta H/c^2 \ll 1$, we have 
\begin{equation}
HT={\rm{constant}}
\end{equation}
Now for the oblique incidence, we can write
\begin{equation}
\frac{\Delta \lambda}{\lambda}=-\frac{2\alpha \Delta H}{c^2}
\cos\theta
\end{equation}
where $\theta$ is the angle of incidence.
This is the change in wavelength per collision. Then following Saha
and Srivastava \cite{R10}, we have the change of wavelength per unit time
\begin{equation}
\frac{\Delta \lambda}{\lambda}= -\frac{\alpha}{c}\frac{\Delta H}{H}
\end{equation}
In the limiting case, we can write
\begin{equation}
\frac{d\lambda}{\lambda}=-\frac{\alpha}{c}\frac{dH}{H}
\end{equation}
Integrating, we have
\begin{equation}
\lambda H^{\omega_c}={\rm{constant}}
\end{equation}
where $\omega_c=\alpha/c$, the frequency of cosmic phonons as
described in \cite{R12}. Then combining eqns.(15) and (19), we have
\begin{equation}
\lambda T^{-\omega_c}={\rm{constant~~ or~~}} \lambda \propto
T^{\omega_c}
\end{equation}
This is the modified version of Wien's displacement law for photon gas in the Rindler space.
It is therefore obvious that $\lambda$ will be quite  large if
$\alpha$ is large enough. This is consistent with the conventional
result on gravitational redshift. Further, the wavelength saturates to a
constant value for $\alpha\longrightarrow 0$.
\section{Gravitational Redshift in Extended Relativistic Dynamics}
Next we consider the prescription of Friedman and Friedman et al
\cite{R7,R8,R9,R91} on
the extended relativistic dynamics and maximal acceleration. Now the
clock hypothesis of Einstein states that the timing of an accelerated
clock is identical with that of a clock at rest in some
un-accelerated frame of reference. It is well known that if the
clock hypothesis is correct, the transformations are Galilean type.
Whereas, if the clock hypothesis is false, then the transformations
are Lorentz type. In the later  case the uniform acceleration of the
frame plays  the role of uniform velocity of Lorentz type
transformations. The maximal value of acceleration plays the role of
velocity of light. In the special theory of relativity the velocity
of light is treated as the upper limit of velocity.

Then following Friedman, 
when the clock hypothesis is false,
we have the proper velocity-time
transformations 
\begin{eqnarray}
t&=& \gamma \left (t^\prime +\frac{\alpha u_x^\prime}{\alpha_m^2}
\right ) \nonumber \\
u_x&=& \gamma (\alpha t^\prime +u_x^\prime )\nonumber \\
u_y&=&u_y^\prime\nonumber \\
u_z&=& u_z^\prime
\end{eqnarray}
where $\gamma=(1- \alpha^2/\alpha_m^2)^{1/2}$, the time dilation
factor. The motion is assumed to be along $x$-direction and the
uniform acceleration $\alpha$ is also along the same direction. Here
$\alpha_m$ is the upper limit of acceleration, just like the velocity
of light is the maximum possible value for velocity in the case of
Lorentz transformation in special theory of relativity.

To get an estimate for Doppler shift of an electromagnetic wave, we
follow Friedman and assume that the wave-vector $k$ in the proper
velocity-time representation is also along $x$-direction. Then from
\cite{R7}, the electromagnetic radiation may be represented by the function
\begin{eqnarray}
(\omega t-k u)&=&f\left [ \omega\gamma \left ( t^\prime
+\frac{\alpha u^\prime}{\alpha_m^2}\right ) -k\gamma (\alpha t^\prime
+u^\prime)\right ] \nonumber \\ &=& f\left [ \gamma \left (\omega
-k\alpha\right )t^\prime -\gamma \left ( k-\frac{\alpha}{\alpha_m^2}
\right )u^\prime \right ] \nonumber \\ &=& f(\omega^\prime t^\prime
-k^\prime u^\prime) ~~ ({\rm{say}})
\end{eqnarray} 
Hence
\begin{eqnarray}
\omega^\prime&=&\gamma(\omega -k \alpha)\nonumber \\ &=& \omega
\frac{\left (1- \frac{\alpha}{\alpha_m}\right )}{ \left
(1-\frac{\alpha^2}{\alpha_m^2}\right )^{1/2}}
\end{eqnarray}
For $\alpha/\alpha_m\ll 1$ and writing $\omega=2\pi \nu$, with $\nu$
the actual frequency, we have
\begin{equation}
\nu^\prime =\nu \left ( 1-\frac{\alpha}{\alpha_m}\right )
\end{equation}
Following the discussion of section-2, here also we consider the outward
expansion of the wall of the enclosure, containing photon gas, with a
uniform acceleration $\alpha$. In this prescription, 
when a photon of frequency $\nu$
incident on the surface of the enclosure, 
it will be red-shifted to $\nu^\prime$ as given
in eqn.(24). Further, if the surface is perfect reflector, the wave
will again be red-shifted without any loss of energy by absorption 
and the final frequency will be
\begin{equation}
\nu^{\prime\prime}=\nu \frac{\left (1-\frac{\alpha}{\alpha_m}\right
)}{\left ( 1+\frac{\alpha}{\alpha_m} \right )}
\end{equation}
Now writing $\nu^{\prime\prime}=\nu+d\nu$, where $d\nu$ is the
effective change in frequency due to Doppler shift, we have
\begin{equation}
\frac{d\nu}{\nu}=-\frac{2\alpha}{\alpha_m}
\end{equation}
Hence
\begin{equation}
\frac{d\lambda}{\lambda}=\frac{2\alpha}{\alpha_m}
\end{equation}
Here we have considered only normal incidence. Now eqn.(27) may be
approximated with  $z$, the gravitational redshift factor. Here the
quantity $z$ depends only on the uniform acceleration $\alpha$.
Further the redshift arises because of the accelerated motion of the 
reflecting surface. 
In accordance with the principle of equivalence, we may replace the
uniformly accelerated frame by a stationary frame in presence of
a constant gravitational field $\alpha$. Therefore one may assume that
$\alpha$ is the constant gravitational field produced by a self
gravitating object, e.g., a neutron star (not a black hole, because
nothing is known about the gravitational field of the black hole). 
Then $\alpha_m$ is the maximum possible surface value of
gravitational field for a neutron star. 
For $\alpha=\alpha_m$, the gravitational redshift factor $z$ becomes $2$.
Since $\alpha$ is always less that $\alpha_m$, therefore 
$z$ is always less than $2$, the bound for gravitational redshift
factor at the neutron star surface for $M/R < 4/9$ \cite{R13}
in geometrical unit, where $M$
and $R$ are respectively the mass and radius of a neutron star.
Therefore according to our
analysis, the quantity $\alpha_m$ has physical significance. It is related to
the maximum gravitational redshift for the compact stellar object concerned.
\section{Wien's Displacement Law for de Broglie Waves in Rindler Space}
To develop a formalism for Wien's displacement law of de Broglie waves in Rindler space
we replace $\hbar \omega$ by the single particle energy $\varepsilon$ of the fermion or boson and further we write
in the non-relativistic approximation the single particle energy $\varepsilon=p^2/2m$, where $m$ is the particle
mass. In this model calculation we assume that the enclosure whose wall is expanding adiabatically with constant 
acceleration is filled up with either a Fermi gas or a Bose gas. At
first we develop the formalism 
in a very general manner, applicable for both the fermions and bosons. At the end of this section only we 
shall differentiate between fermions from bosons. Further, the
collisions of the particles are assumed to be elastic in nature.

Now following eqns.(1) and (3), we have
\begin{equation}
p_1=p_0\left ( 1+\frac{\alpha \Delta H}{c^2}\right )^{1/2}\approx p_0\left (1+\frac{\alpha \Delta H}{2c^2} \right )
\end{equation}
and
\begin{equation}
p_1\approx p_2\left ( 1-\frac{\alpha \Delta H}{2c^2}\right )
\end{equation}
Combining these two equations, we have
\begin{equation}
P_2=p_0\frac{1+\frac{\alpha \Delta H}{2c^2}} {1- \frac{\alpha \Delta
H}{2c^2}}
\end{equation}
Now writing $\lambda=h/p$, the de Broglie wave length of the
particle, we have
\begin{equation}
\lambda_2=\lambda_0\frac{1-\frac{\alpha \Delta H} {2c^2}} {1+ \frac{\alpha \Delta
H}{2c^2}}
\end{equation}
This is the relation between the initial (before incidence) and final (after
reflection) de Broglie wavelengths of the particle.
Now as before, assuming an infinitesimal change in de Broglie
wavelength during reflection from the wall, we can write
$\lambda_2=\lambda+\delta \lambda$, where we have
put $\lambda$ for $\lambda_0$, the initial de Broglie wavelength and
$\delta \lambda$ is the infinitesimal change in de Broglie wavelength.
Then following the same procedure as has been used for photon gas, we
have in the limiting case
\begin{equation}
\frac{d\lambda}{\lambda}=-\omega_c \frac{dH}{H}
\end{equation}
On integrating, we have
\begin{equation}
\lambda H^{\omega_c}={\rm{constant}}
\end{equation}
Now for a non-relativistic gas
\begin{equation}
P=\frac{2}{3}\epsilon
\end{equation}
where $\epsilon$ is the energy density and $P$ is the kinetic
pressure of the gas. Now for the adiabatic expansion of
non-relativistic gas, we have 
\begin{equation}
PV^{5/3}=\rm{constant}
\end{equation}
Combining these two equations, we can write
\begin{equation}
\epsilon V^{5/3}={\rm{constant}}
\end{equation}
Now for a non-relativistic gas obeying Boltzman statistics, we can
write
\begin{equation}
\epsilon = C T^{5/2}\exp\left (\frac{\mu}{kT}\right )
\end{equation}
where $C$ is a constant and $\mu$ is the chemical potential of the
gas. Combining eqns.(36) and (37) and using $V=44\pi H^3/3$ as the volume
of the enclosure, with $H$ its radius, we have
\begin{equation}
TH^2\exp\left (\frac{2\mu}{5kT}\right ) ={\rm{constant}}
\end{equation}
Again using eqn.(33), we have for a Fermi gas with non-zero chemical
potential
\begin{equation}
\lambda ={\rm{constant}} T^{0.5{\omega_c}}\exp \left (\frac{\mu
\omega_c}{5kT} \right )
\end{equation}
This is the modified form of Wien's displacement law for fermionic de
Broglie wavelength in Rindler space. Obviously the wavelength
increases in a much faster rate with the strength of gravitational
field $\alpha$ in comparison with photon gas for fixed $T$. Further,
the wavelength exponentially diverges for large $\alpha$. Similar to
the photon gas the de Broglie wavelength saturates to a constant
value for very low $\alpha$ ($\alpha \rightarrow 0$). On the other
hand if one considers an anti-fermion with chemical potential $-\mu$,
then the de Broglie wavelength after reflection becomes
\begin{equation}
\lambda ={\rm{constant}} T^{0.5{\omega_c}}\exp \left (\frac{-\mu
\omega_c}{5kT} \right )
\end{equation}
Therefore as the gravitational field becomes strong enough the de
Broglie wavelength of the anti-fermion decreases and in the extreme
case for $\alpha\rightarrow \infty$, $\lambda \rightarrow 0$.
Therefore  in extremely strong gravitational field anti-fermions
behave like classical objects. Since anti-particles do not exist in
classical mechanics, even in the non-relativistic quantum mechanics
they do not exist. In Schr\"{o}dinger equation particle and the
corresponding anti-particle are treated in equal footing. Therefore
we can conclude that Strong gravitational field will never allow the
emission of anti-particles. When the pairs are produced at the surface of a 
strongly gravitating object, then we expect only the particle will be
emitted, whereas the anti-particle counter part will go inside the 
object. Which may be the another version of Penrose mechanism, but the reason is quite different.

Next we consider a Bose gas with chemical potential $\mu=0$ (say a
mixture of $\pi^+$ and $\pi^-$.) Then it is obvious that
the modified form of Wien's displacement law is given by
\begin{equation}
\lambda={\rm{constant}} T^{0.5{\omega_c}}
\end{equation}
The variation is therefore little slower compared to photon gas.
\section{Conclusion}
In this article we have developed a formalism to study Wien's
displacement law for photon gas and a gas of de Broglie waves
in a frame undergoing a uniform accelerated motion.

It is well known that the conventional form of Wien's displacement
law for photon gas 
is $\lambda \propto T^{-1}$, whereas in the present scenario
it is $\lambda \propto T^{\omega_c}$, with $\omega_c=\alpha/c
\geq 0$. Hence one can infer that for $\alpha \longrightarrow
\infty$, $\lambda \longrightarrow \infty$. This is consistent with the
results of gravitational redshift in presence of strong gravitational
field produced by some compact massive stellar objects. 
In this case
the redshift is because of uniform acceleration of the frame or
because of constant gravitational field in the sense of principle of
equivalence. Therefore not the temperature, but the strength of
gravitational field plays the major role in producing Doppler shift
of the wave if the field is extremely strong.
It is also obvious that for low gravitational field
(acceleration), i.e., for $\alpha\longrightarrow 0$, the wavelength
$\lambda$ saturates to some constant value. This is true for the
formalism either based on the prescription of Feynman lecture or using
Rindler coordinates. 

In the second part we have studied the
gravitational redshift for electromagnetic waves in the context of
extended relativistic dynamics with an upper limit of acceleration
$\alpha_m$ of the frame. In this scenario we have noticed that for
$\alpha=\alpha_m$, the gravitational redshift factor $z=2$. Since
$\alpha$ is always $ < \alpha_m$, we have $z < 2$, the bound of
gravitational redshift factor. This can also be obtained from general theory
of relativity with $M/R < 4/9$ \cite{R14}. 

In the third part we have studied the gravitational redshift of de
Broglie waves in Rindler space. In the case of fermionic de Broglie
waves the effect is more prominent because of the exponential term.
The wavelength increases much faster with the gravitational field
compared to photon gas. For extremely high gravitational field the
wavelength diverges exponentially. Whereas for a Bose gas with zero
chemical potential the increase of wavelength with gravitational
filed is little bit slower compared to photon gas. For extremely high
field it again diverges, but not exponentially. 

The physical reason for the gravitational redshift of photons is the
curvature of the space produced by gravity, whereas in the case of de
Broglie waves it is because of the gain or loss of energy of the
particle, when it goes in favor or against gravity respectively. 


\begin{thebibliography}{99}
\bibitem{R61} The fractional change of $d\nu/\nu$ as  given by
eqn.(7) was experimentally verified by Pound and Rebeka and later
with more accurately by Pound and Snider. They confirm the prediction
using the $74$ft. high Jefferson tower and shown that the fractional
change is $\approx 2.5 \times 10^{-15}$. The ratio of experimental to
theoretical values is $\sim 0.999 \pm 0.0076$. For the detail
discussion, see two references as given below.
\end{thebibliography}
\end{document}